%% file: KrLoMuSampledFunctObsv.tex
\documentclass[journal,twoside,web]{ieeecolor}
\usepackage{lcsys}
\usepackage{cite}
\usepackage{amsmath,amssymb,amsfonts}
\usepackage{algorithmic}
\usepackage{graphicx}
\usepackage{textcomp}
\usepackage{microtype}
\usepackage{tikz}

\usepackage{graphics} 
\usepackage{epsfig} 
\usepackage{mathptmx} 
\usepackage{times} 
\usepackage{amsmath} 
\usepackage{amssymb}  
\usepackage{tcolorbox}
\usepackage{pgfkeys}
\usepackage{xcolor}
\usepackage{pgfplots}
\usepackage{layouts}
\usepackage{tikz}
\usepackage[ngerman]{babel}
\usepackage{mathtools}
\usepackage{microtype}

\newlength\figurewidth
\definecolor{rev}{rgb}{0,0,0}
\definecolor{rew}{rgb}{0,0,0} 
\definecolor{fin}{rgb}{0,0,0} 

\newcommand\copyrighttext{%
	\footnotesize \copyright 2025 IEEE. Personal use of this material is permitted. Permission from IEEE must be obtained for all other uses, in any current or future media, including reprinting/republishing this material for advertising or promotional purposes, creating new collective works, for resale or redistribution to servers or lists, or reuse of any copyrighted component of this work in other works.}
\newcommand\copyrightnotice{%
	\begin{tikzpicture}[remember picture,overlay]
		\node[anchor=south,yshift=7pt] at (current page.south) {\fbox{\parbox{\dimexpr\textwidth-\fboxsep-\fboxrule\relax}{\copyrighttext}}};
	\end{tikzpicture}%
}

\def\BibTeX{{\rm B\kern-.05em{\sc i\kern-.025em b}\kern-.08em
    T\kern-.1667em\lower.7ex\hbox{E}\kern-.125emX}}
\begin{document}

	\newtheorem{thm}{Theorem}
	\newtheorem{cor}{Corollary}
	\newtheorem{lem}{Lemma}
	\newtheorem{prop}{Proposition}
	
	\newtheorem{rem}{Remark}
	
	\newtheorem{defi}{Definition}
	\newtheorem{ass}{Assumption}
	
	\setlength\figurewidth{0.9\columnwidth}

\title{	On sample-based  functional observability  of linear systems}
\author{Isabelle Krauss,  Victor G. Lopez, \IEEEmembership{Member, IEEE}, and Matthias A. Müller, \IEEEmembership{Senior Member, IEEE}
	\thanks{This work received funding from the European Research Council (ERC) under the European Union's Horizon 2020 research and innovation programme (grant agreement No 948679).}
	\thanks{I. Krauss,  V. G. Lopez and M. A. M\"uller are with the Leibniz University Hannover, Institute of Automatic Control,  30167 Hannover,  Germany
		{\tt\small \{krauss,lopez,mueller\}@irt.uni-hannover.de}}
}

\maketitle
\thispagestyle{empty}
\copyrightnotice

\begin{abstract}
	Sample-based observability characterizes the ability to reconstruct the internal state of a dynamical system by using limited
	output information, i.e., when measurements are only infrequently and/or irregularly available. In this work, we investigate the concept of functional observability, which refers to the ability to infer a function of the system state from the outputs, within a sample-based framework. Here, we give necessary and sufficient conditions for a system to be sample-based functionally observable, and formulate conditions on the sampling schemes such that these are satisfied. Furthermore, we provide a numerical example, where we demonstrate the applicability of the obtained results.
\end{abstract}

\section{INTRODUCTION}
\IEEEPARstart{I}{n} various engineering applications, such as state-feedback control and system monitoring, it is necessary to estimate the internal state of a system.
For designing conventional state estimators, state observability is typically needed.
However, some systems require a more nuanced approach to observability.
For example, in many practical applications, complete state estimation is not necessary for fulfilling the system's operational requirements. Instead, reconstructing specific functions or combinations of state variables is sufficient and often more efficient.
\textcolor{rew}{This is relevant, for example,  in high-dimensional or networked systems, where only a subset of variables is required for control, monitoring, or diagnosis - as seen in applications ranging from power grids and supply networks to biomedical systems \cite{Mon22}.}
This is where the concept of functional observability, first proposed in \cite{FTJ10}, becomes relevant. Functional observability extends the classical notion of observability to the ability to reconstruct a function of the state from the system's external outputs, even when the full state itself is not observable \cite{FJT10,JFT11}. 
Algebraic criteria to characterize functional observability have been studied in  \mbox{\cite{FJT10,JFT11,FZ15,ZFD24}}.
\par In some situations/applications, the system's output measurements are sparse or irregular, often due to practical constraints that prevent taking measurements continuously or at every time instant. 
Such instances of limited output information can occur, for example, in the biomedical field when analyzing blood samples from a patient that are collected only occasionally. These measurements are used, for instance, to assess hormone levels for diagnosing disorders in the hypothalamic–pituitary–thyroid axis \cite{Die16} and to formulate effective medication strategies (as discussed in \cite{Bru21, Wol22}). Irregular sampling sequences may also be produced by event-triggered  sampling in e.g. networked control systems to decrease computation and communication loads (see \cite{Ge20, Mis15,Pen18}). 
In these scenarios, it is essential to employ state estimators capable of handling limited output information to accurately reconstruct the internal state. To design suitable state estimators,  sample-based observability or detectability conditions,
that take this limited output information into account, are needed.
For linear systems, such sample-based observability conditions were investigated in \cite{WLY11,ZA16} for continuous-time systems and in \cite{Kra22} for discrete-time systems. 
The case of nonlinear discrete-time systems was also addressed in \cite{KLM25}.
Although sample-based observability has been studied, other concepts like  sample-based functional observability remain absent in the current literature. In this paper we now aim  to explore partially observable and  functionally observable linear systems within a sample-based framework.  
\par In particular, our contributions are as follows. We extend the results on sample-based observability to partially observable systems (Section~\ref{sec:PO}).  Then, we introduce the more general concept of sample-based functional observability and provide necessary and sufficient conditions for a system to be sample-based functionally observable (Section~\ref{sec:FO}). Moreover, we discuss the corresponding sampling strategies (Section \ref{sec:Sampstart}) and provide conditions for these to be relaxed (Section~\ref{sec:Relax}).
Finally, we illustrate how to use these results to reconstruct a function of the state (Section \ref{sec:NumEx}). 
\section{Preliminaries and Setup}
\label{sec:Pre}
Consider a linear time-invariant system described by
\begin{align}
	\begin{aligned}
		x^+&=Ax(t)+Bu(t),\\
		y(t)&=Cx(t)+Du(t), \\	&z(t)=Fx(t) ,
	\end{aligned}
	\label{eq:sys2}
\end{align}
where $A \in \mathbb{R}^{n \times n}$, $B \in \mathbb{R}^{n \times m}$, $C \in \mathbb{R}^{q \times n}$, $D \in \mathbb{R}^{q \times m}$, $F \in \mathbb{R}^{r \times n}$, $u(t) \in \mathbb{R}^m$ is the control input, $x(t) \in \mathbb{R}^n$ is the state and $y(t) \in \mathbb{R}^q$ is the system output. Additionally, in Section~\ref{sec:FO} we consider $z(t)\in \mathbb{R}^r$ as a function of the state aimed to be reconstructed.
Throughout the paper we study both continuous-time and discrete-time systems, thus we use the notation $x^+$  to denote the time derivative $\dot{x}(t)$ in the context of continuous-time systems and the state $x(t+1)$ when referring to discrete-time systems.
$O(A,C)$ refers to the observability matrix
, i.e., $O(A,C)=\begin{pmatrix}C^\top&(CA)^\top&(CA^2)^\top&\cdots&(CA^{(n-1)})^\top\end{pmatrix}^\top$.
	\par In the following,  we introduce the concept of sample-based observability and review some results on the conditions for observable linear continuous-time and discrete-time systems to be sample-based observable. 
	 Considering explicitly the (potentially irregular) time instances of a measurement sequence in the formulation of the observability matrix leads to 
	what we call sample-based observability matrix. \textcolor{rew}{These matrices have been used to study conditions on sampling strategies for sample-based observability, cf. \cite{ZA16,Kra22}, where it was also discussed that full column rank of these matrices allow the unique reconstruction of the initial state $x_0$ based on the available output samples.}
	\begin{defi}[Sample-based observability matrix]
		\label{def:SOm}
		Consider a set of arbitrary time instances \color{rew} $\{ t_i \}_{i=1}^k$ for some $k\geq 1$, with $0 \leq t_1$ and $t_i < t_{i+1}$ for all $i=1,...,k-1$.
		\color{black}
		For continuous-time systems the sample-based observability matrix  is given by
		\begin{align}
			O_{\text{s}}(A,C)=\begin{pmatrix}
				(Ce^{At_1})^\top&(Ce^{At_2})^\top& \ldots&(Ce^{At_k})^\top 
			\end{pmatrix}^\top.
			\label{eq:Osc}
		\end{align}
		For discrete-time systems it  is formulated as follows
		\begin{align}
			O_{\text{s}}(A,C)=\begin{pmatrix}
				(CA^{t_1})^\top&(CA^{t_2})^\top& \ldots&(CA^{t_k})^\top 
			\end{pmatrix}^\top.
			\label{eq:Osd}
		\end{align}
	\end{defi}
	\par In the subsequent sections, we frequently address both discrete-time and continuous-time systems simultaneously. For simplification, we uniformly label the sample-based observability matrices for both continuous-time systems and discrete-time as  $O_{\text{s}}$.
	If we specifically focus only on either  continuous or discrete-time systems, $O_{\text{s}}$ refers to (\ref{eq:Osc}) or (\ref{eq:Osd}), respectively.
	\begin{defi}[Sample-based Observability]
		\label{def:SO}
		System (\ref{eq:sys2}) is  sample-based  observable for a given sampling sequence 	\textcolor{rew}{$\{ t_i \}_{i=1}^k$ for some $k\geq 1$, with $0 \leq t_1$ and $t_i < t_{i+1}$ for all $i=1,...,k-1$},  if for any initial state $x_0$ and input $u(t)$, 
		the value of $x_0$ can be uniquely reconstructed 	\textcolor{rew}{from the knowledge of the input trajectory on the interval $[0,t_k]$ and the sampled  outputs $y(t_i)$, $i=1,\ldots,k$,} 
		i.e., the sample-based observability matrix $O_{\text{s}}$ in Definition \ref{def:SOm} has full column rank.
	\end{defi}
	\par For observable continuous-time systems, some sampling strategies to guarantee full column rank of (\ref{eq:Osc}) are provided in \cite{WLY11,ZA16}. 
	\textcolor{rew}{There, the number of time instances where the signal $t \rightarrow Ce^{At} \Delta x_0$ becomes zero for $\Delta x_0 \neq 0$ is studied, where $\Delta x_0$  refers to the difference of two different initial states.} An upper bound on the number of zeros of this signal inside a fixed time interval can be computed by 
	\begin{equation}
		k^{*}= d-1+\frac{T\delta}{2\pi}, 
		\label{eq:sbObsvC}
	\end{equation}
	where $T$ is the length of the considered time interval, $\delta=\max_{1\leq i,j\leq v} \Im (\lambda_i-\lambda_j)$ with  $\Im(\cdot)$ denoting the imaginary part, $d=\sum_{i=1}^v d_i$, $d_i$ is the index of the eigenvalue $\lambda_i$ of the matrix $A$, and $v$ is the number of pairwise distinct eigenvalues of the system \cite{ZA16}. Selecting $k>k^*$ measurements in a $T$-long time interval renders (\ref{eq:Osc}) full column rank.
	\par  In \cite{Kra22} it was observed that, in contrast to continuous-time systems, deriving a (meaningful) bound on \emph{arbitrary} samples in order
	to guarantee sample-based observability of a discrete-time system is not possible in
	general. 
	Nevertheless, a variety of sampling strategies were devised for different classes of system dynamics to guarantee full column rank of (\ref{eq:Osd}). These conditions for sample-based observability are, as in the continuous-time case (see \textcolor{rew}{equation} (\ref{eq:sbObsvC})), formulated in terms of the system's eigenvalues.
	\par In the following, we investigate how to leverage these results on sampling strategies in the context of functionally observable systems. For the clarity of our arguments, it will be convenient to discuss in the following section the case of partially observable systems,  i.e., systems where only parts of the internal state can be reconstructed from output measurements. Later, we investigate the more general case of functionally observable systems.
	\section{Extension to partially observable systems}
	\label{sec:PO}
		In the following, it will be useful to consider system~(\ref{eq:sys2}) in some canonical decomposition  $(A_\text{o},B_\text{o},C_\text{o},D_\text{o})$ which separates the observable part of the system, that we later use for the design of the sampling scheme, i.e., 
			\begin{align}
				A_{\text{o}}=P_{\text{o}}^{-1}AP_{\text{o}}=\begin{pmatrix} A_{\text{ob}} & 0\\A_{21}&A_{\text{$\overline{\text{ob}}$}}\end{pmatrix}, \ C_{\text{o}}=CP_{\text{o}}= \begin{pmatrix} C_{\text{ob}} & 0\end{pmatrix}\label{eq:Ao}\end{align} 
			\textcolor{rew}{for} some invertible $P_{\text{o}}\in \mathbb{R}^{n\times n}$ 
			\cite[Chapter 6.4]{Che99}.
				Here, $(A_{\text{ob}},C_{\text{ob}})$ corresponds to the observable subsystem of (\ref{eq:sys2}).
				We use $p$ to denote the dimension of the unobservable subspace of (\ref{eq:sys2}), thus $A_{\text{ob}}\in \mathbb{R}^{(n-p)\times(n-p)}$,  $C_{\text{ob}}\in \mathbb{R}^{q\times(n-p)}$. 
				\textcolor{rew}{Applying the sampling strategies discussed in Section II to  the system $(A_{\text{ob}},C_{\text{ob}})$  such that $O_{\text{s}}(A_{\text{ob}},C_{\text{ob}})$ has full column rank}
				also makes the observable subspace, \textcolor{rew}{ i.e., the orthogonal complement of the unobservable subspace,} of the original system sample-based observable.
				This is formulated in the following theorem. The proof is performed for continuous-time systems. The discrete-time case follows in a similar fashion. 
			\begin{thm}
				Consider the pair $(A_{\text{ob}},C_{\text{ob}})$ and a sampling scheme such that 
				\begin{align} 
					\mathtt{rank}(	O_{\text{s}}(A_{\text{ob}},C_{\text{ob}})	)=n-p.
					\label{eq:PO}
				\end{align}
				Then, for such a sampling scheme it holds that  $O_{\text{s}}(A,C)h\neq0$ for all $h\notin \mathtt{null}(O(A,C))$ \textcolor{rew}{with $h \in \mathbb{R}^n$.}
				\label{thm:POc}
			\end{thm}  
			\begin{proof}	
				Using (\ref{eq:Ao}) and (\ref{eq:PO}), it holds that $\mathtt{rank}(	O_{\text{s}}(A_{\text{o}},C_{\text{o}})	)=\mathtt{rank}(	O_{\text{s}}(A_{\text{ob}},C_{\text{ob}})	)=n-p$ 
				which is the maximum possible rank.
				Furthermore, we can write
				\begin{align*}
					\begin{aligned}
						&O_{\text{s}}(A_{\text{o}},C_{\text{o}})	 \\&=\begin{pmatrix}
							CP_{\text{o}}e^{P_{\text{o}}^{-1}AP_{\text{o}}t_1} 
							\\ \vdots \\ CP_{\text{o}}e^{P_{\text{o}}^{-1}AP_{\text{o}}t_k}
						\end{pmatrix}
						=\begin{pmatrix}
							CP_{\text{o}}P_{\text{o}}^{-1}e^{At_1}P_{\text{o}}
							\\ \vdots \\ CP_{\text{o}}P_{\text{o}}^{-1}e^{At_k}P_{\text{o}}
						\end{pmatrix}
						=\begin{pmatrix}
							Ce^{At_1}P_{\text{o}}
							\\ \vdots \\ Ce^{At_k}P_{\text{o}}
						\end{pmatrix}\\
						&	=O_{\text{s}}(A,C)P_{\text{o}}.
					\end{aligned}
				\end{align*}
				Hence, $\mathtt{rank}(O_{\text{s}}(A_{\text{o}},C_{\text{o}})	)
				=\mathtt{rank}(O_{\text{s}}(A,C))=n-p$. 
				Furthermore, since  $e^{At_i}$ can be expressed as a linear combination of $I,A,A^2,...,A^{n-1}$ by the Cayley-Hamilton theorem, we have $\mathtt{rank}\begin{pmatrix}
					O(A,C)\\O_{\text{s}}(A,C)
				\end{pmatrix}=n-p$. 
				Combining the above, it follows that $\mathtt{null}(O_{\text{s}}(A,C))= \mathtt{null}(O(A,C))$ and thus $O_{\text{s}}(A,C)h\neq0$ for all $h\notin \mathtt{null}(O(A,C))$.
			\end{proof} 
			\section{Sample-based functional observability}
			\label{sec:FO}
			We now turn to the more general case of functional observability, where only some function (linear combination) of the states shall be reconstructed. 
			Note that a system does not necessarily require observability in order to be functionally observable.
			In the following we define sample-based functional observability, provide necessary and sufficient conditions for a functionally observable triple $(A,C,F)$ to be sample-based functionally observable and relate these findings to the  results from Section \ref{sec:PO}. Unless specified otherwise, explanations and statements apply to both continuous-time and discrete-time systems.
			\begin{defi}[Functional Observability \cite{FJT10}]
				The triple $(A,C,F)$ is functionally observable, if for
				any initial state $x_0$ and input $u(t)$, there exists a finite time $\bar{t}$
				such that the value of $F x_0$ can be uniquely reconstructed from
				the outputs $y(t)$ and inputs $u(t)$, $0\leq t \leq \bar{t}$.
				\label{def:FO}
			\end{defi}
			\begin{lem}[{\cite{ZFD24}}]
				The following three statements are equivalent.
				\begin{enumerate}
					\item $(A,C,F)$ is functionally observable. 
					\item $\mathtt{rank}\begin{pmatrix}O(A,C)\\O(A,F)\end{pmatrix}=\mathtt{rank}(O(A,C))$.
					\item $\mathtt{rank}\begin{pmatrix}O(A,C)\\F\end{pmatrix}=\mathtt{rank}(O(A,C))$.
				\end{enumerate}
				\label{lem:FO}
			\end{lem}
			\par Next we define sample-based functional observability.	
			\textcolor{rew}{Here, we will use the observable subsystem of $(A,F)$.  For this, let $(A_{\text{o,F}},F_{\text{o}})$  denote the canonical decomposition of $(A,F)$  analogous to (\ref{eq:Ao}), i.e., $A_{\text{o,F}}=P_{\text{o,F}}^{-1}AP_{\text{o,F}}=\begin{pmatrix} A_{\text{ob,F}} & 0\\A_{21\text{,F}}&A_{\text{$\overline{\text{ob}}$,F}}\end{pmatrix}, \ F_{\text{o}}=FP_{\text{o,F}}= \begin{pmatrix} F_{\text{ob}} & 0\end{pmatrix}$. Thus, $(A_{\text{ob,F}},F_{\text{ob}})$ refers to the observable subsystem of $(A,F)$. We denote $\nu$ as the observability   index of $(A_{\text{ob,F}},F_{\text{ob}})$. Using this notation, we define $\sigma=\begin{cases}0,&\text{(\ref{eq:sys2}) is a continuous-time system} \\\nu-2, &\text{(\ref{eq:sys2}) is a discrete-time system.}\end{cases}$}
			\color{rew}
			\begin{defi}[Sample-based Functional Observability]
				The triple $(A,C,F)$ is functionally observable  for a given sampling sequence  $\{ t_i \}_{i=1}^k$ for some $k\geq 1$, with $0 \leq t_1$ and $t_i < t_{i+1}$ for all $i=1,...,k-1$,  if there \textcolor{fin}{exists} a finite time $\bar{t}>\sigma$ such that for any initial state $x_0$ and input $u(t)$, 
				the value of $F x(t),$ $0 \leq t \leq \bar{t}$ can be uniquely reconstructed from the knowledge of the input trajectory on the interval $[0,\bar{t}]$ and the sampled  outputs $y(t_i)$, $i=1,\ldots,k$.
				\label{def:FOs}
			\end{defi}
			\color{black} \par Now let us review the three statements in Lemma~\ref{lem:FO} for the sample-based case.
			An obvious adaptation to the sample-based setting would be to replace all observability matrices by their sample-based counterparts, i.e.,
			\begin{enumerate}
				\item[(i)] $(A,C,F)$ is sample-based functionally observable. \\
				\item[(ii)] $\mathtt{rank}\begin{pmatrix}O_{\text{s}}(A,C)\\O_{\text{s}}(A,F)\end{pmatrix}=\mathtt{rank}(O_{\text{s}}(A,C))$
				\\
				\item[(iii)] $\mathtt{rank}\begin{pmatrix}O_{\text{s}}(A,C)\\F\end{pmatrix}=\mathtt{rank}(O_{\text{s}}(A,C))$
			\end{enumerate}
			It turns out that these three conditions are \emph{not} equivalent. In fact it holds that  
			\textcolor{rew}{(i)	$\Rightarrow$ (ii) and (i) 	$\Rightarrow$ (iii)}. 
			\textcolor{fin}{The facts that the implications (i)	$\Rightarrow$ (ii) and (i) 	$\Rightarrow$ (iii) hold, whereas the converse implications do not, are all direct consequences of Theorem~\ref{lem1} below.} 
			\color{rew}	In the following, we demonstrate that (ii) and  (iii) do not necessarily imply each other by providing some counter examples.
			To  show that (iii) $\nRightarrow$ (ii) we
			consider the following triple of a discrete-time system
			\color{black}
			\begin{align*}
				\begin{aligned}
					A=\begin{pmatrix} 1& 1 & 0& 0 \\-1 &1& 0& 0\\ 0& 0  &2&2\\0 &0 &-2& 2\end{pmatrix},\ \begin{array}{c} C=\begin{pmatrix} 1 &1& 1 &1\end{pmatrix},\\[0.4em]
						F=\begin{pmatrix} 1& 1 &0 &0\end{pmatrix}.\end{array}
				\end{aligned}
			\end{align*}
			This system is observable and thus also functionally observable. The system has a pathological period of 4, meaning that taking every fourth measurement results in $\mathtt{rank}(O_{\text{s}}(A,C))=2$ (cf. \cite{Kra22} for further details on pathological sampling periods).  Now consider the following set of measurement  instances $\{0,4,8,13\}$. This sampling sequence yields $\mathtt{rank}(O_{\text{s}}(A,C))=3$ and $\mathtt{rank}\begin{pmatrix}O_{\text{s}}(A,C)\\F\end{pmatrix}=3$. However, the vector $FA^{13}$ is linearly independent of the rows of the matrix $\begin{pmatrix}O_{\text{s}}(A,C)\\F\end{pmatrix}$. Hence, $\mathtt{rank}\begin{pmatrix}O_{\text{s}}(A,C)\\O_{\text{s}}(A,F)\end{pmatrix}=4$ and thus (iii) $\nRightarrow$ (ii). 
			\color{rew}
			To see that (ii) $\nRightarrow$ (iii), consider  a pathological sampling sequence 
			$\{2,6,10,14\}$. Then,  due to the pathological sampling sequence and the system being functionally observable, $O_{\text{s}}(A,F)$ does not increase the rank, i.e.,  $\mathtt{rank}\begin{pmatrix}O_{\text{s}}(A,C)\\O_{\text{s}}(A,F)\end{pmatrix}=2$. However, 
			$F$ is linearly independent of the rows of $O_{\text{s}}(A,C)$ and thus increases the rank, i.e., $\mathtt{rank}\begin{pmatrix}O_{\text{s}}(A,C)\\F\end{pmatrix}=3$.
			In a similar fashion we can construct  counter examples in continuous time. 	
			\par \color{black} Statements (ii) and (iii)  as stated above are both only necessary conditions for sample-based functional observability. In the following theorem we establish a  necessary and sufficient condition for sample-based functional observability.	
			\begin{thm}
				\label{lem1}
				The triple $(A,C,F)$ is sample-based functionally observable if and only if  \begin{align}\mathtt{rank}\begin{pmatrix}O_{\text{s}}(A,C)\\O(A,F)\end{pmatrix}=\mathtt{rank}(O_{\text{s}}(A,C))\label{eq:thmFO}.
				\end{align}
			\end{thm}
			\begin{proof}
				By Definition \ref{def:FOs} we know that $(A,C,F)$ is sample-based functionally observable if and only if for any initial state $x_0$ and zero input, $y(t)=Cx(t)=0$ for all $t \in \{t_1,\ldots,t_k\}$ implies that $z(t)=F x(t)=0$ for all $t\in [0,\textcolor{rew}{\bar{t}}]$ \textcolor{rew}{ with $\bar{t}$ as in Definition~\ref{def:FOs}.} 
				\par 	Sufficiency part: If   (\ref{eq:thmFO}) holds, then \begin{align*}\begin{pmatrix} y(t_1)^\top \ldots y(t_k)^\top\end{pmatrix}^\top=O_s(A,C)x_0=0 
				\end{align*} 
				implies $Fx(t)=0$ for all $t\in[0,\textcolor{rew}{\bar{t}}]$. 
				\par \color{black}	Necessity part:
				This we show by contradiction. Let $(A,C,F)$ be sample-based functionally observable. Assume 	
				$\mathtt{rank}\begin{pmatrix}O_s(A,C)\\O(A,F)\end{pmatrix}>\mathtt{rank}(O_s(A,C))$.
			Then, $\mathtt{null}\begin{pmatrix}O_s(A,C)\\O(A,F)\end{pmatrix} \subset  \mathtt{null}\begin{pmatrix}O_s(A,C)\end{pmatrix}$ \textcolor{rew}{where $ \subset $ denotes the strict inclusion}. This implies that there exists an $x_0\neq0$ that is in the nullspace of  $O_s(A,C)$ and thus yields $y(t)=0$ for all $t \in \{t_1,\ldots,t_k\}$, but is not in the nullspace of  $\begin{pmatrix}O_s(A,C)\\O(A,F)\end{pmatrix} $ and thus results in $z(t)\neq0$ for some $t\in[0,\textcolor{rew}{\bar{t}}]$. 
	\end{proof} 
	\subsection{Sampling strategies for sample-based functional observability}
	\label{sec:Sampstart}
	Now, we relate the results from Section \ref{sec:PO} to functional observability.
	In particular, a sampling scheme that results in sample-based observability of the observable subsystem 
	also guarantees sample-based functional observability.
	\begin{thm}
		If $(A,C,F)$ is functionally observable, then applying sampling schemes such that \textcolor{rew}{$\mathtt{rank}(O_{\text{s}}(A_{\text{ob}},C_{\text{ob}}))=n-p$} guarantees
		sample-based functional observability of $(A,C,F)$.
		\label{thm:thm12}
	\end{thm}
	\begin{proof}
		As discussed in the proof of Theorem~\ref{thm:POc},
		by applying the proposed sampling scheme, it holds that $\mathtt{rank} (O_{\text{s}}(A,C))=n-p$, i.e., the maximum possible rank. Hence,  
		\begin{align}
			\mathtt{null}(O_{\text{s}}(A,C))= \mathtt{null}(O(A,C)).
			\label{eq:rankcontra}
		\end{align}
		Since $(A,C,F)$ is functionally observable, by Lemma~\ref{lem:FO} it holds that	
		\begin{align}
			\mathtt{rank}\begin{pmatrix}
				O(A,C)\\	O(A,F)
			\end{pmatrix}=\mathtt{rank}(O(A,C)).
			\label{eq:FO_lem}
		\end{align}
		Combining (\ref{eq:rankcontra}) and (\ref{eq:FO_lem}) we conclude that 
		\begin{align*}
			\mathtt{rank}\begin{pmatrix}O_{\text{s}}(A,C)\\O(A,F)\end{pmatrix}=\mathtt{rank}(O_{\text{s}}(A,C)),
		\end{align*}
		i.e., sample-based functional observability of $(A,C,F)$ (cf. Theorem \ref{lem1}).
		\end{proof}
		\par	
		\begin{rem}
			Theorem~\ref{thm:thm12}	shows that considering the eigenvalues of  $A_{\text{ob}}$ in the design of the sampling scheme guarantees sample-based functional observability of any functionally observable triple  $(A,C,F)$. 
			In fact, consider the case where $(A,\tilde{C},F)$ is a functionally observable triple, where the notation $\tilde{C}\in \mathbb{R}^{\tilde{q}\times n}$ refers to a submatrix of $C$. Then, from the application of Theorem~\ref{lem1} with $\tilde{C}$ instead of $C$ and following analogous steps as in the proof of Theorem~\ref{lem:FO}, it is sufficient to
			consider the eigenvalues of  $(\tilde{A}_{\text{ob}},\tilde{C}_{\text{ob}})$, i.e., the observable subsystem of $(A,\tilde{C})$, to design a sampling scheme that guarantees sample-based functional observability of $(A,C,F)$.
		\end{rem}
		\subsection{Relaxing the sampling schemes}
		\label{sec:Relax}
		From the previous result (Theorem~\ref{thm:thm12}) we know that sample-based functional observability is achieved when considering the eigenvalues of $A_{\text{ob}}$ in the design of the sampling scheme. 
		However, these sampling strategies to guarantee sample-based functional observability can be relaxed for certain matrices $F$. 
		For these cases it is sufficient to consider only the part of the observable subsystem that influences $z=Fx$, i.e., only the eigenvalues of $(A_{\text{ob,F}},F_{\text{ob}})$,  need to be considered in the design of the sampling scheme. 
		\color{rew} The order of the system $(A_{\text{ob,F}},F_{\text{ob}})$ is denoted by $n-p_F$ with $p_F$ referring to the dimension of the unobservable subspace of $(A,F)$.
		\color{black}
		\par First we address the case when $F$ can be expressed as a linear combination of the rows of $C$.
		\begin{lem}
			If $\mathtt{rank}\begin{pmatrix}
				C\\F
			\end{pmatrix}= \mathtt{rank}(C)$,
			then using a sampling scheme such that 
			\begin{align}
				\mathtt{rank}(O_{\text{s}}(A_{\text{ob,F}},F_{\text{ob}}))=n-p_F
				\label{eq:lemCF}
			\end{align}
			%
			guarantees sample-based functional observability. 
			\label{lem:CF}
		\end{lem}
		\begin{proof}
			By  Theorem~\ref{thm:POc}, the applied sampling strategy implies $\mathtt{rank}\begin{pmatrix}O_{\text{s}}(A_{\text{ob,F}},F_{\text{ob}})\end{pmatrix}=\mathtt{rank}\begin{pmatrix}O_{\text{s}}(A,F)\end{pmatrix}=\mathtt{rank}(O(A,F))$ \textcolor{rew}{and that $\mathtt{null}(O_{\text{s}}(A,F)) = \mathtt{null}(O(A,F))$.}
			Furthermore,  due to $F$ belonging to the row space of $C$,  we know that $F=\alpha C$ with $\alpha \in \mathbb{R}^{r\times q}$. Therefore, since 
			\begin{align*}
				\mathtt{rank}\begin{pmatrix}O_{\text{s}}(A,C)\\O_{\text{s}}(A,\alpha C)\end{pmatrix}=\mathtt{rank}(O_{\text{s}}(A,C)),
			\end{align*}
			we can conclude 
			\begin{align*}
				\mathtt{rank}\begin{pmatrix}O_{\text{s}}(A,C)\\O(A,F)\end{pmatrix}=\mathtt{rank}(O_{\text{s}}(A,C)),
			\end{align*}
			and the result then follows from Theorem~\ref{lem1}.
		\end{proof}
		\color{rew} \par For a specific class of systems, it is possible to obtain a similar result but with a matrix $F$ that does not lie within the row space of $C$. In particular, in the following theorem we consider a slightly  more general condition on $F$ for systems where the eigenvalues of $A$ have a geometric multiplicity of one. 
		This condition is formulated in terms of a matrix $Q$  of a specific structure.  To this end, denote the Jordan canonical form\footnote{\color{rew} Note that we consider the Jordan canonical form, and not the real Jordan form, i.e., if $A$ has complex eigenvalues, they appear on the diagonal of $A_{\text{J}}$.} of system  (\ref{eq:sys2})  by $(A_{\text{J}},B_{\text{J}},C_{\text{J}},D_{\text{J}},F_{\text{J}})$.
		\begin{defi}
			We define $Q\in \mathbb{C}^{n\times n}$ as a nonsingular block diagonal matrix,  i.e., $ Q=\mathtt{diag}\{Q_j\}_{j=1}^w$,  where each block $Q_j$  has the same dimension $k_j$ as the corresponding Jordan block of $A_{\text{J}}$.  Moreover, each $Q_j$ is given by $Q_j = q_{j,1} I + q_{j,2} U_j + q_{j,3} U_j^2 + \cdots + q_{j,k_j} U_j^{k_j-1}$, for given constants $q_{j,i}$, and where $U_j$ is the upper shift matrix\footnote{\color{rew} An upper shift matrix has ones on the superdiagonal and zeros everywhere else.} of dimension $k_j$. 
			\label{def:Q}
		\end{defi}
		\par Thus, for example, if $A$ has two Jordan blocks with dimensions 3 and 1, respectively, then Q is given by 
		\color{black}
		\begin{align*}
			\textcolor{rew}{	Q=\begin{pmatrix}
					q_{1,1} &q_{1,2} &q_{1,3} & 0\\0& q_{1,1} &q_{1,2}& 0\\0 &0 &q_{1,1}& 0\\0& 0 &0& q_{2,1} 
				\end{pmatrix}.}
		\end{align*}
		\begin{thm}
			Consider system (\ref{eq:sys2}) in its Jordan canonical form. 
			We assume that the system does not exhibit multiple Jordan blocks associated
			with the same eigenvalue.	
			Suppose $F_{\text{J}}=\alpha C_{\text{J}}Q$ for some  $\alpha \in \mathbb{R}^{r\times q}$ and \textcolor{rew}{some $Q$ as in Definition~\ref{def:Q}.} Then, using a sampling scheme such that (\ref{eq:lemCF}) holds  guarantees sample-based functional observability. 
			\label{lem:lem4}
		\end{thm}
		\begin{proof}				
			If $A_{\text{J}}$  does not have  multiple Jordan blocks associated with the same eigenvalue, then $(A_{\text{J}}, \alpha C_{\text{J}})$ and  $(A_{\text{J}}, \alpha C_{\text{J}} Q)$  have the same unobservable subspace. 	In the following, for clarity we first show this for a system composed of a single Jordan block of dimension  three with eigenvalue \textcolor{rew}{$\lambda \in \mathbb{C}$} and then discuss how the corresponding argument can be extended.
			The corresponding  observability matrix 	$O(A_{\text{J}},\alpha C_{\text{J}})$ with \mbox{$\alpha C_{\text{J}}=\begin{pmatrix} c_1&c_2&c_3\end{pmatrix}$} can be expressed as 
			\begin{align*}
				\underbrace{
					\begin{pmatrix}
						1& 0 & 0\\
						\lambda &1 & 0
						\\
						\lambda^2& 2\lambda &1 
				\end{pmatrix}}_{\text{$\Gamma$}} 
				\underbrace{
					\begin{pmatrix}
						c_1&c_2&c_3\\ 0&c_1 &c_2\\ 0 & 0&c_1
				\end{pmatrix}}_{\text{$\Psi_1$}}.
			\end{align*}
			It is clear that $\mathtt{null}(O(A_{\text{J}},\alpha C_{\text{J}}))=\mathtt{null}(\Psi_1)$. Analogously, $O(A_{\text{J}},\alpha C_{\text{J}}Q)$ can be expressed as follows
			\color{rew}	
			\begin{align*}
				\Gamma
				\underbrace{
					\begin{pmatrix}
						c_1q_{1,1}&c_1q_{1,2}+c_2q_{1,1}&c_1q_{1,3}+c_2q_{1,2}+c_3q_{1,1}\\ 0&c_1q_{1,1}&c_1q_{1,2}+c_2q_{1,1}\\ 0 & 0&c_1q_{1,1}
					\end{pmatrix}
				}_{\text{$\Psi_2$}} 
			\end{align*}
			Since $Q$ is invertible and thus all diagonal elements $q_{1,1}$ are nonzero, \color{black} $\mathtt{null}(\Psi_1)=\mathtt{null}(\Psi_2)$, and it follows	
			\begin{align}
				\mathtt{null}(	O(A_{\text{J}},\alpha C_{\text{J}}))=\mathtt{null}(	O(A_{\text{J}},\alpha C_{\text{J}}Q)).
				\label{eq:null}
			\end{align}
			\textcolor{rew}{This can be easily extended to systems with $w>1$ Jordan blocks of arbitrary dimensions, where the matrix $\Gamma$ takes the form $\Gamma = \begin{pmatrix}\Gamma_1, \ldots, \Gamma_w\end{pmatrix}$. Since the eigenvalues have geometric multiplicity of one, all the submatrices $\Gamma_i$ are different and $\Gamma$ has full column rank, making (\ref{eq:null}) hold.
			}
			Next, we show that 
			\begin{align}
				\mathtt{rank}(O_{\text{s}}(A_{\text{J}},\alpha C_{\text{J}}))=	\mathtt{rank}(O_{\text{s}}(A_{\text{J}},\alpha C_{\text{J}} Q)).
				\label{eq:Rankeq}
			\end{align}
			\color{rew}
			For this, first  note that due to the structure of $Q$ as in Definition~\ref{def:Q},  
			it follows that $A_{\text{J}}=Q^{-1}A_{\text{J}} Q$. 
			Therefore, 
			\begin{align*}
				\begin{aligned}
					&	O_{\text{s}}(A_{\text{J}},\alpha C_{\text{J}})Q\\&
					=\begin{pmatrix}
						\alpha C_{\text{J}} Q Q^{-1} A_{\text{J}}^{t_1} Q \\ \vdots \\ \alpha C_{\text{J}} Q Q^{-1} A_{\text{J}}^{t_k}Q
					\end{pmatrix}=\begin{pmatrix}
						\alpha C_{\text{J}}  Q A_{\text{J}}^{t_1} \\ \vdots \\ \alpha C_{\text{J}}  Q A_{\text{J}}^{t_k} 
					\end{pmatrix}=O_{\text{s}}(A_{\text{J}},\alpha C_{\text{J}}Q).
				\end{aligned}
			\end{align*}
			and thus (\ref{eq:Rankeq}) follows.
			\par From (\ref{eq:lemCF}) and Theorem~\ref{thm:POc}  it can be shown that  
			$\mathtt{null}( O_{\text{s}}(A_{\text{J}},F_{\text{J}}))=\mathtt{null}(O(A_{\text{J}},F_{\text{J}})). 
			$ This fact, together with (\ref{eq:null}) and (\ref{eq:Rankeq}), implies that $\mathtt{null}(O_{\text{s}}(A_{\text{J}},\alpha C_{\text{J}}))=\mathtt{null}(O(A_{\text{J}},F_{\text{J}}))$.
			Combining this and the fact that 
			\begin{align*}
				\mathtt{rank}\begin{pmatrix}O_{\text{s}}(A_{\text{J}},C_{\text{J}})\\O_{\text{s}}(A_{\text{J}},\alpha C_{\text{J}})\end{pmatrix}=\mathtt{rank}(O_{\text{s}}(A_{\text{J}},C_{\text{J}}))
			\end{align*} we can conclude 
			\begin{align*}
				\mathtt{rank}\begin{pmatrix}O_{\text{s}}(A_{\text{J}},C_{\text{J}})\\O(A_{\text{J}}, F_{\text{J}})\end{pmatrix}=\mathtt{rank}(O_{\text{s}}(A_{\text{J}},C_{\text{J}}))
			\end{align*}
			which is equivalent to (\ref{eq:thmFO})
			and the result again follows from Theorem~\ref{lem1}.
		\end{proof}
		\color{black}
		\par Theorem~\ref{lem:lem4} implies that in the design of the measurement sequence it is sufficient to only consider the modes that are observable through $F$, i.e., the modes that appear in the signal $Fe^{At}$ (continuous-time systems) or $FA^t$ (discrete-time systems),  when a linear combination of the rows of $C$ can produce a matrix such that $(A,\alpha C)$ has the same observable subspace as $(A,F)$.
			\section{Application to least squares state estimation}
			\label{sec:NumEx} 
			For illustration purposes we now design a least squares estimator to show how the results from the previous sections can be applied to reconstruct $z$ (cf. \cite{WLY11,ZIA17} where least squares estimation was addressed in the context of continuous-time systems with sampled outputs). 
			Subsequently, a numerical example of the aforementioned is provided, also illustrating the significance of Theorem~\ref{lem:lem4}.
			\subsection{Estimator design }
			\label{sec:Est}
			For simplicity and without loss of generality, let us consider here autonomous systems. 
			\textcolor{rev}{
				The aim is now to estimate the observable part of the state  in the canonical decomposition $(A_{\text{o}},C_{\text{o}})$, i.e., ${x}_{\text{ob}}$, by  minimizing
				\begin{align*}
					\min_{\hat{x}_{\text{ob}}(t_{j})}	\textcolor{rew}{||\Phi_{j,k} \hat{x}_{\text{ob}}(t_{j})-Y_{j,k}||^2}
				\end{align*}
				\textcolor{rew}{where $||\cdot||$ refers to the Euclidean norm. } Here,
				\begin{align*}
					\begin{aligned}
						Y_{j,k}&=\begin{pmatrix}
							y(t_{j})\\
							y(t_{j+1}) \\\vdots\\y(t_{j+k-1})
						\end{pmatrix}, \quad 
						\Phi_{j,k}&=\begin{pmatrix}
							C_{\text{ob}}\\C_{\text{ob}}A_{\text{ob}}^{t_{j+1}-t_{j}}\\\vdots\\C_{\text{ob}}A_{\text{ob}}^{t_{j+k-1}-t_{j}}
						\end{pmatrix}
					\end{aligned}
					\label{eq:lsmatrices}
				\end{align*}
				in the discrete-time case, whereas for continuous-time systems \textcolor{rew}{ terms of the form $C_{\text{ob}}A_{\text{ob}}^{t}$ in $\Phi_{j,k}$ are replaced by $C_{\text{ob}}e^{A_{\text{ob}}t}$.} 
				The \textcolor{rew}{indices  $j,k$ of $\Phi$ and $Y$ refer} to the time of the first measurement ($t_j$) and the number of measurements ($k$)  considered in the minimization problem.
				By designing the measurement sequences as discussed in Theorem~\ref{thm:POc} and Theorem~\ref{thm:thm12}, that is considering the observable modes in the design of the sampling strategy, it is guaranteed that $\Phi$ has full column rank and thus the left inverse $(\Phi^\top \Phi)^{-1}\Phi^\top$ exists.}
			When the goal is to estimate a function of the states, then we can obtain $\hat{z}({t_{j}})$ by 
			$	\hat{z}(t_{j})=(FP_{\text{o}})_{\text{ob}}(\Phi_{j,k}^\top \Phi_{j,k})^{-1}\Phi_{j,k}^\top Y_{j,k}$,
			where $(FP_{\text{o}})_{\text{ob}}$ refers to the submatrix of $FP_{\text{o}}$ corresponding to $A_{\text{ob}}$, i.e., the first $n-p$ columns of $FP_{\text{o}}$.
			This least squares approach can now be used to reconstruct the initial state of a system or to consistently estimate the current state based on the last $k$ measurements. In the latter case, the estimate of the current function of the state is obtained by  $\hat{z}(t_{j+k-1})=(FP_{\text{o}})_{\text{ob}}A_{\text{ob}}^{t_{j+k-1}-t_{j}}(\Phi_{j,k}^\top \Phi_{j,k})^{-1}\Phi_{j,k}^\top Y_{j,k}$ where $\hat{x}_{\text{ob}}(t_{j})=(\Phi_{j,k}^\top \Phi_{j,k})^{-1}\Phi_{j,k}^\top Y_{j,k}$. Until the next measurement is available  at $t_{j+k}$,  $\hat{z}(\tau)$ for $\tau \in(t_{k+j-1},t_{k+j})$ is obtained by running the observer in open loop, i.e., $\hat{z}(\tau)=(FP_{\text{o}})_{\text{ob}}A_{\text{ob}}^{\tau-t_{j+k-1}}\hat{x}(t_{j+k-1})$ or $\hat{z}(\tau)=(FP_{\text{o}})_{\text{ob}}e^{A_{\text{ob}}(\tau-t_{j+k-1})}\hat{x}(t_{j+k-1})$ for discrete-time and continuous-time systems, respectively.
			\par \textcolor{rev}{
				If Lemma~\ref{lem:CF}  or
				Theorem~\ref{lem:lem4} is applicable, we can instead set up the possibly lower dimensional matrices $\tilde{\Phi}$ and $\tilde{Y}$ as follows
				\begin{align*}
					\begin{aligned}
						\tilde{Y}_{j,k}&=\begin{pmatrix}
							\alpha y(t_{j})
							\\\vdots\\\alpha y(t_{j+k-1})
						\end{pmatrix}, \ 
						\tilde{\Phi}_{j,k}&=\begin{pmatrix}
							(\alpha	CP_{\text{o,F}})_{\text{ob}}
							\\\vdots\\(\alpha CP_{\text{o,F}})_{\text{ob}}A_{\text{ob,F}}^{t_{j+k-1}-t_{j}}
						\end{pmatrix}
							\end{aligned}
						\end{align*}
						where $\alpha$ is such that $F_{\text{J}}=\alpha C_{\text{J}}Q$ and $(\alpha	CP_{\text{o,F}})_{\text{ob}}$ refers to the submatrix of $\alpha	CP_{\text{o,F}}$ consisting of the first $n-p_F$ columns of $\alpha	CP_{\text{o,F}}$.
						The continuous-time version of $\tilde{\Phi}$ is formulated analogously. The minimization problem is accordingly adapted to
						\begin{align*}
							\min_{\hat{x}_{\text{ob,F}}(t_{j})}	\textcolor{rew}{||\tilde{\Phi}_{j,k} \hat{x}_{\text{ob,F}}(t_{j})-\tilde{Y}_{j,k}||^2}
						\end{align*}
						with $\hat{x}_{\text{ob},F}$ being the state estimate of the system described by the pair $(A_{\text{ob,F}},(\alpha CP_{\text{o,F}})_{\text{ob}})$. 
						Adapting the matrices as described above guarantees that $\tilde{\Phi}$ has still full column rank even if $\Phi$ does not. }
								\subsection{Simulation results}
								\par Consider the following discrete-time system with additive measurement noise $d(t)$
								\begin{align*}
									x(t+1)&=\begin{pmatrix}
										1& 0 &0&0\\1&-1&0 &0\\-2&3.2&0.6&2.4\\0&-0.3&-0.15&0.6
									\end{pmatrix}x(t),\\
									y(t)&=\begin{pmatrix}
										1&0&0&0\\1.5&-4&-1&0\\2&-6&-1&-4
									\end{pmatrix}x(t)+d(t),\\
									z(t)&=\begin{pmatrix}
										0&-2&-1&1
									\end{pmatrix}x(t).
								\end{align*}
								\textcolor{rew}{
									Note that the property of sample-based (functional) observability is not affected by measurement noise (as is the case for standard observability). However, in this case, only an approximate reconstruction of the state (or functional output) is possible, which can readily be obtained by the procedure described in \textcolor{fin}{Section~\ref{sec:Est}}.
							}
							\par
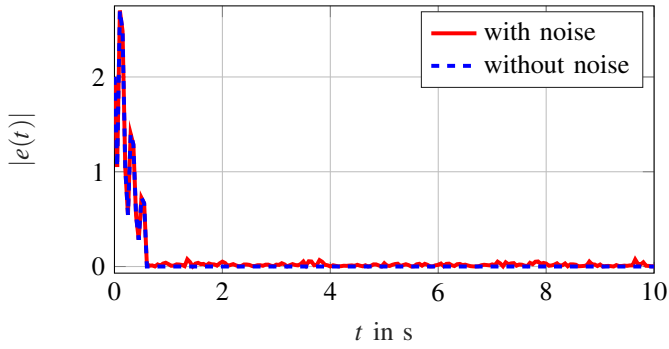
\begin{figure}[!t]
								\centering
								\centerline{\input{figures/error_z}}
								\caption{Estimation error of $z(t)$ for the nominal case (blue dashed) and for the case of noisy measurements (red solid).} 
								\label{fig:error}
							\end{figure}

 Here, we can apply Theorem~\ref{lem:lem4}. It holds that $F_{\text{J}}=\alpha C_{\text{J}}Q=\begin{pmatrix}0&0&\textcolor{rew}{1-4i}&\textcolor{rew}{1+4i}\end{pmatrix}$ for $	\alpha= \begin{pmatrix}1&-2&1 \end{pmatrix}$ and $Q=\mathtt{diag}\{1,1,\textcolor{rew}{-0.625
								+0.375
								i,-0.625
								-0.375
								i}\}$.
							Therefore, we only need to consider $\lambda_3$ and $\lambda_4$, that refer to the observable eigenvalues of $(A,F)$, 
							in the design of the measurement sequence, i.e., such that (\ref{eq:lemCF}) holds. Hence, since $A_{\text{ob,F}}$ is a $2\times2$ matrix, we can use a result on second-order systems (cf. \cite[Lemma 3]{Kra22}). Therefore, and since we do not need to consider $\lambda_1$ and $\lambda_2$, fewer measurements are required to estimate $z$ than if the aim would be to reconstruct the complete state $x$.
							Now we obtain $\hat{z}(t)$ as described above.
							Figure~\ref{fig:error} shows the estimation error for the nominal case and when considering noisy measurements, \textcolor{fin}{ where $d$ is modeled as a uniformly distributed random variable satisfying $|d(t)|\leq 0.1$.} 
							The estimation error is initially large due to the estimator running  in open loop  based on a prior estimate  differing from the true initial state until enough measurements are collected. In the nominal case, $z$ is then accurately reconstructed, whereas in case with noisy measurements, the estimation error converges to a small neighborhood of zero. 
							\section{Conclusion}
							\label{sec:con}
							In this paper we studied observability conditions that are of relevance when only parts (or functions) of the state need to be reconstructed from irregular or infrequent measurement sequences, for both linear continuous-time and discrete-time systems. 
							First we extended the results on sample-based observability to only partially observable systems. Then, we proposed the concept of sample-based functional observability, provided  necessary and sufficient conditions for sample-based functional observability, and derived conditions on the sampling schemes such that these are satisfied. In the general case, this is given if the observable modes of the system can be reconstructed, which can be relaxed to the observable modes of $(A,F)$ for special cases of the functional output matrix $F$.
							We illustrated the obtained results by using them for (functional) state estimation via a least squares estimator in a numerical example. Using the derived sample-based functional observability conditions in a more sophisticated (functional) estimator such as, e.g., moving horizon estimator, is an interesting subject of future work.

\end{document}

%% file: figures/error_z.tex
%
%
\begin{tikzpicture}

\begin{axis}[%
width=0.9\figurewidth,
height=1.4in,
scale only axis,
xmin=0,
xmax=10,
xlabel style={font=\color{white!15!black}},
xlabel={$t$ in  s},
ymin=-0.07,
ymax=2.75,
ylabel style={font=\color{white!15!black}},
ylabel={$|e(t)|$},
axis background/.style={fill=white},
xmajorgrids,
ymajorgrids,
legend style={legend cell align=left, align=left, draw=white!15!black}
]
\addplot [color=red, line width=1.5pt]
  table[row sep=crcr]{%
0	2\\
0.05	1.05\\
0.1	2.7\\
0.15	2.484\\
0.2	1.0368\\
0.25	0.54432\\
0.3	1.39968\\
0.35	1.2877056\\
0.4	0.53747712\\
0.45	0.282175488\\
0.5	0.725594112\\
0.55	0.66754658304\\
0.6	0.0268338090078892\\
0.65	0.00826245386595881\\
0.7	0.0116239170111132\\
0.75	0.00059846507962294\\
0.8	0.0220293411475348\\
0.85	0.00864447895134834\\
0.9	0.0312373173002131\\
0.95	0.0400210409222017\\
1	0.0255343806504885\\
1.05	0.00809205375241437\\
1.1	0.00989800259341231\\
1.15	0.0228350896810544\\
1.2	0.0211818822287567\\
1.25	0.0165472687295109\\
1.3	0.00342470369793969\\
1.35	0.0765219310291485\\
1.4	0.0466011246060369\\
1.45	0.000825559186257013\\
1.5	0.0308391783674875\\
1.55	0.0388808245170806\\
1.6	0.0244527809959054\\
1.65	0.0274616533185751\\
1.7	0.0105492115000686\\
1.75	0.0324314441894561\\
1.8	0.0235818356839356\\
1.85	0.0221448587664849\\
1.9	0.00959490882734839\\
1.95	0.0338242639148035\\
2	0.0513009632607442\\
2.05	0.0372076858942348\\
2.1	0.00678500170723217\\
2.15	0.016449537006917\\
2.2	0.0252658493571476\\
2.25	0.0124956832333625\\
2.3	0.00319659165711112\\
2.35	0.00493105641651857\\
2.4	0.0213935725260945\\
2.45	0.0213208093130403\\
2.5	0.0173225866296426\\
2.55	0.0054361212501819\\
2.6	0.00504312772308259\\
2.65	0.0134390452549778\\
2.7	0.0140057928704492\\
2.75	0.00783245916748547\\
2.8	0.0194831218677059\\
2.85	0.0199384402946456\\
2.9	0.00865066923004429\\
2.95	0.00397487393609155\\
3	0.0200192067941068\\
3.05	0.0170831172962544\\
3.1	0.0349135696472618\\
3.15	0.0227504012226915\\
3.2	0.00116817568706094\\
3.25	0.0141758156554868\\
3.3	0.0254441950269529\\
3.35	0.0158758656054631\\
3.4	0.0121750126884699\\
3.45	0.0260406384620972\\
3.5	0.0532704438255348\\
3.55	0.00881996587330814\\
3.6	0.055365729605106\\
3.65	0.053228377465706\\
3.7	0.0128773821162416\\
3.75	0.0167340753690498\\
3.8	0.0677763543786917\\
3.85	0.0497251312964736\\
3.9	0.0178216969965413\\
3.95	0.0129981715769182\\
4	4.54400029387386e-06\\
4.05	0.0132387081853927\\
4.1	0.00730051799115565\\
4.15	0.00168990362543288\\
4.2	0.00385681977993046\\
4.25	0.00823080705613191\\
4.3	0.00710005822580808\\
4.35	0.00994245970834025\\
4.4	0.0117502029700815\\
4.45	0.00694167257409264\\
4.5	0.018135145240175\\
4.55	0.00890363670529117\\
4.6	0.00237294052657646\\
4.65	0.0129684848936597\\
4.7	5.389355962867e-05\\
4.75	0.00696601086815915\\
4.8	0.0226132189753435\\
4.85	0.0221203349453378\\
4.9	0.010805807701373\\
4.95	0.00077342106153667\\
5	0.034855525288081\\
5.05	0.00902566642205363\\
5.1	0.00600171589141079\\
5.15	0.0183448779447871\\
5.2	0.0263350889755604\\
5.25	0.0296247920891054\\
5.3	0.00477966961850549\\
5.35	0.0155942467619494\\
5.4	0.0240943586621626\\
5.45	0.0148869533034399\\
5.5	0.000516405727370614\\
5.55	0.00308912276390517\\
5.6	0.0233104098437461\\
5.65	0.0287956165606226\\
5.7	0.00915535423599344\\
5.75	0.0283992231104165\\
5.8	0.0181443536609866\\
5.85	0.00132578375368413\\
5.9	0.00193954093442727\\
5.95	0.0132251682254607\\
6	0.0172666713433404\\
6.05	0.0314107124407276\\
6.1	0.00440271627122341\\
6.15	0.0278989724827922\\
6.2	0.0302335107398452\\
6.25	0.0140128610125947\\
6.3	0.00495269451757488\\
6.35	0.0243117565032005\\
6.4	0.0275068968622023\\
6.45	0.0155038115523383\\
6.5	0.0194365569027554\\
6.55	0.0110707718626115\\
6.6	0.00375150125119031\\
6.65	0.00123866337233258\\
6.7	0.00121468485405808\\
6.75	0.00300552193162645\\
6.8	0.0136471596769519\\
6.85	0.0185405674031133\\
6.9	0.0153139763286985\\
6.95	0.0120907863005753\\
7	0.0224857066866155\\
7.05	0.0151698609111153\\
7.1	0.0510098504977343\\
7.15	0.0321314962221049\\
7.2	0.00183070310815749\\
7.25	0.0448208553604953\\
7.3	0.024854965880181\\
7.35	0.0330464373577621\\
7.4	0.0358593124416696\\
7.45	0.00881013352088642\\
7.5	0.0213384939993426\\
7.55	0.0192628966641734\\
7.6	0.00949975134886561\\
7.65	0.00430750674785444\\
7.7	0.0389317745311715\\
7.75	0.0245322194190988\\
7.8	0.050672684099047\\
7.85	0.0532104631270591\\
7.9	0.0273682232011567\\
7.95	0.0379993064681858\\
8	0.0311010422806704\\
8.05	0.00817173683296216\\
8.1	0.0113178036995619\\
8.15	0.019465014959207\\
8.2	0.0337522581208782\\
8.25	0.0092866386421498\\
8.3	0.0131576594764523\\
8.35	0.0322035675924411\\
8.4	0.00236219074322927\\
8.45	0.022910477867893\\
8.5	0.0226946623233424\\
8.55	0.0187159679433632\\
8.6	0.00259913004030652\\
8.65	0.00430228064620701\\
8.7	0.00633220738786836\\
8.75	0.0106962909307107\\
8.8	0.0522969601738063\\
8.85	0.0223669211134803\\
8.9	0.0209950680570239\\
8.95	0.0049478421421267\\
9	0.0210538595716088\\
9.05	0.000524811384239778\\
9.1	0.00483007121987406\\
9.15	0.0066686746670993\\
9.2	0.00402194767428746\\
9.25	2.48914488336849e-05\\
9.3	0.0152106402075143\\
9.35	0.00550268871609034\\
9.4	0.00584440148457007\\
9.45	0.00945480101402973\\
9.5	0.00713779214794508\\
9.55	0.0116752462754805\\
9.6	0.0119746755969941\\
9.65	0.0744822660503482\\
9.7	0.028205399207678\\
9.75	0.0213204580618542\\
9.8	0.044715659646839\\
9.85	0.0079549058279363\\
9.9	0.00629236589003038\\
9.95	0.00228068034904436\\
10	0.00278920256965551\\
};
\addlegendentry{with noise}

\addplot [color=blue, dashed, line width=1.5pt]
  table[row sep=crcr]{%
0	2\\
0.05	1.05\\
0.1	2.7\\
0.15	2.484\\
0.2	1.0368\\
0.25	0.54432\\
0.3	1.39968\\
0.35	1.2877056\\
0.4	0.53747712\\
0.45	0.282175488\\
0.5	0.725594112\\
0.55	0.66754658304\\
0.6	2.77555756156289e-16\\
0.65	3.33066907387547e-16\\
0.7	2.77555756156289e-16\\
0.75	2.77555756156289e-16\\
0.8	1.94289029309402e-16\\
0.85	3.33066907387547e-16\\
0.9	1.66533453693773e-16\\
0.95	1.94289029309402e-16\\
1	5.55111512312578e-17\\
1.05	8.32667268468867e-17\\
1.1	3.33066907387547e-16\\
1.15	8.32667268468867e-17\\
1.2	1.2490009027033e-16\\
1.25	3.60822483003176e-16\\
1.3	2.22044604925031e-16\\
1.35	6.93889390390723e-17\\
1.4	1.52655665885959e-16\\
1.45	6.93889390390723e-17\\
1.5	2.4980018054066e-16\\
1.55	6.48786580015326e-16\\
1.6	1.6306400674182e-16\\
1.65	1.28369537222284e-16\\
1.7	2.18575157973078e-16\\
1.75	8.23993651088983e-17\\
1.8	6.07153216591882e-17\\
1.85	2.56739074444567e-16\\
1.9	2.27248775352962e-16\\
1.95	3.5301622736128e-16\\
2	1.00613961606655e-16\\
2.05	7.11236625150491e-17\\
2.1	6.15826833971767e-17\\
2.15	2.93385107874577e-16\\
2.2	5.44269490587723e-17\\
2.25	4.42354486374086e-17\\
2.3	7.97972798949331e-17\\
2.35	7.2099444470286e-17\\
2.4	5.41016884070267e-17\\
2.45	9.15066633577766e-17\\
2.5	1.27285335049798e-16\\
2.55	3.15340201867409e-16\\
2.6	1.43765208071578e-16\\
2.65	3.83807569059869e-16\\
2.7	4.32054565735474e-16\\
2.75	2.43186547288499e-16\\
2.8	1.23978518423717e-16\\
2.85	8.04478011984244e-17\\
2.9	1.23599047663348e-16\\
2.95	1.10724146865082e-17\\
3	1.97758476261356e-16\\
3.05	4.60054086839912e-16\\
3.1	3.04389759925305e-17\\
3.15	1.52330405234213e-16\\
3.2	1.38168014356121e-16\\
3.25	2.64992563482613e-16\\
3.3	4.78133158066107e-17\\
3.35	3.51078215977962e-17\\
3.4	2.30298093963077e-16\\
3.45	3.6949610038306e-16\\
3.5	2.3825342740369e-17\\
3.55	1.40468555840864e-16\\
3.6	2.79995211044382e-17\\
3.65	1.65882932390282e-17\\
3.7	2.07014852308951e-16\\
3.75	4.38873180960871e-16\\
3.8	1.99064601066022e-16\\
3.85	7.93331058398378e-17\\
3.9	2.17865344363279e-16\\
3.95	1.87243409254141e-16\\
4	3.54229178541748e-18\\
4.05	2.24346840475668e-16\\
4.1	2.15026089924082e-16\\
4.15	2.50425290855734e-16\\
4.2	1.9304643197578e-16\\
4.25	1.0931807217264e-17\\
4.3	1.28552919855364e-16\\
4.35	1.98056420100553e-16\\
4.4	1.61648614686958e-16\\
4.45	2.01213523653206e-16\\
4.5	6.57363212122749e-17\\
4.55	3.13685946521421e-17\\
4.6	1.92442073967914e-16\\
4.65	2.2085050022764e-16\\
4.7	2.81442578593002e-16\\
4.75	2.36828929744645e-16\\
4.8	1.96013164873539e-16\\
4.85	1.61066279535721e-16\\
4.9	1.1305888730507e-16\\
4.95	7.98274554436791e-17\\
5	2.57323315419936e-18\\
5.05	3.0334478596619e-16\\
5.1	2.31086152241379e-16\\
5.15	3.6080228656134e-17\\
5.2	1.77264593511877e-16\\
5.25	5.21948319543e-17\\
5.3	1.03081209763329e-16\\
5.35	2.03267916039988e-16\\
5.4	9.9948451790464e-17\\
5.45	2.21236456084012e-16\\
5.5	1.0474389573412e-16\\
5.55	3.69971984006907e-16\\
5.6	6.86016152971545e-17\\
5.65	3.90870500350989e-17\\
5.7	4.53891803034628e-17\\
5.75	1.91206072572497e-16\\
5.8	2.16347062620744e-16\\
5.85	2.35546124815808e-16\\
5.9	8.61831605459568e-17\\
5.95	6.61660200338003e-18\\
6	1.27723393358594e-16\\
6.05	2.87808448090647e-16\\
6.1	8.94660683436124e-17\\
6.15	1.57264849848773e-16\\
6.2	6.87511927129613e-17\\
6.25	3.02725945680159e-16\\
6.3	4.37640989559354e-16\\
6.35	1.67741660166248e-16\\
6.4	1.13087033779774e-17\\
6.45	1.20422707198368e-17\\
6.5	2.1880668417217e-16\\
6.55	2.08881143100655e-16\\
6.6	3.70859636485321e-17\\
6.65	2.79124365635243e-16\\
6.7	1.53704299705695e-16\\
6.75	2.72580944507451e-16\\
6.8	3.38963462419577e-16\\
6.85	2.06678707614705e-16\\
6.9	8.68685785315402e-18\\
6.95	5.83034869521345e-17\\
7	2.92333531514909e-16\\
7.05	6.99531774947597e-17\\
7.1	2.13700769255928e-16\\
7.15	4.48824427031812e-16\\
7.2	1.25780007317547e-17\\
7.25	8.858418376976e-17\\
7.3	7.1295560572505e-17\\
7.35	2.23206733107613e-16\\
7.4	3.24672363440169e-16\\
7.45	2.18365085607901e-16\\
7.5	2.68128871869764e-16\\
7.55	4.63564165782433e-18\\
7.6	1.12525065870244e-16\\
7.65	2.29844790444173e-16\\
7.7	1.21210381257738e-16\\
7.75	2.49904699858151e-16\\
7.8	2.88378211750935e-16\\
7.85	2.53727297452402e-16\\
7.9	2.25568321865731e-16\\
7.95	6.16346553247e-17\\
8	7.13730149151687e-17\\
8.05	2.07080349804328e-16\\
8.1	4.19999986078599e-16\\
8.15	3.09782667719009e-16\\
8.2	1.95217814458531e-17\\
8.25	4.08139147233386e-17\\
8.3	3.48447997182487e-16\\
8.35	4.68372040058748e-16\\
8.4	5.12527909749379e-16\\
8.45	1.95440949722949e-16\\
8.5	1.97139165817255e-16\\
8.55	1.43358809751155e-16\\
8.6	2.54860342066429e-17\\
8.65	1.05633032251293e-17\\
8.7	1.16737642214419e-16\\
8.75	5.01517725720702e-16\\
8.8	3.14649563727743e-16\\
8.85	3.20096167082331e-16\\
8.9	7.74109929250446e-17\\
8.95	1.43609068307138e-17\\
9	4.35069052965731e-16\\
9.05	6.38586330392946e-17\\
9.1	8.09132264216506e-17\\
9.15	2.46960998984159e-16\\
9.2	5.45661945334356e-18\\
9.25	5.9484777513545e-17\\
9.3	1.15323423391865e-16\\
9.35	2.85226563000088e-16\\
9.4	4.01565680237213e-16\\
9.45	3.79682009431761e-16\\
9.5	2.17180391361999e-17\\
9.55	7.04851479713868e-17\\
9.6	6.53653271633254e-17\\
9.65	1.29527726307415e-16\\
9.7	3.63423164671855e-17\\
9.75	2.808955717577e-16\\
9.8	4.29825497210646e-16\\
9.85	1.52599551464453e-16\\
9.9	3.65865216826607e-16\\
9.95	1.47868945306364e-16\\
10	9.00472727983693e-17\\
};
\addlegendentry{without noise}

\end{axis}

\end{tikzpicture}%